\documentstyle[aps,pre]{revtex}
\begin{document}

\title{Number of bonds in the site-diluted lattices: sampling and
fluctuations}
\author{Lev N. Shchur and Oleg A. Vasilyev}
\address{Landau Institute for Theoretical Physics, 142432 Chernogolovka,
Russia}
\maketitle
\begin{abstract}

We have calculated analytically the mean value and the variance of the
number of bonds on the lattices of dimension $d$ for the given occupation
of sites. We consider both kinds of site occupation: with the fixed
concentration $n_s$ of occupied sites and with the probability $p$ for a
site to be occupied. We found that the variance are smaller in the former
case and not depends from the dimensionality of the hypercubic lattice.
Whereas in the last case it grows with the space dimensionality. The ratio
of variances are quite different in the limit of $p\rightarrow 1$.
Finally, we demonstrate the relevance of the level of fluctuations on the
precision of energy calculations for the Ising model in the Monte Carlo
simulations.

\end{abstract}


\draft

\section{Introduction}

There are two ways to generate on the lattice a sample with the randomly
occupied sites~\cite{RD-12} . The first one, with the fixed number $s$ of
occupied sites. This is the sampling from a `canonical ensemble' for which
in this paper we will use the name {\it s-sampling}. The second way, with
a fixed probability $p$ for a given site to be occupied, i.e. sampling
from a `grand canonical ensemble', or {\it p-sampling}. It is known that
both methods lead in the 'thermodynamic' limit of infinite lattice size to
the same results for the free energy per Ising spin being placed onto
occupied sites of the lattice~\cite{GrifLeb}.

If one choose the probability $p$ for the site to be occupied the
concentration of bonds between the occupied sites on the infinite lattice
are $n_{SS}=p^2$. We named this bonds as $SS$-bonds. Similarly, the
concentration of bonds between non-occupied sites ($NN$-bonds) is
$n_{NN}=(1-p)^2$ and the concentration of bonds between occupied and
non-occupied sites ($SN$-bonds) are $n_{SN}=2p(1-p)$. Clearly, the sum of
the three concentrations $n_{SS}+n_{NN}+n_{SN}$ is the probability that
each bond belongs to one of the three types and equal to unity.

In this paper we asking the following questions: i) how fluctuates the
concentration of bonds on the hypercubic lattices depending on the
concentration of occupied sites? ii) how this fluctuations depend on the
way of sampling? iii) how this fluctuations depend on the lattice size?

We found analytically the answers on the all of the stated questions.

We start our analyses with the simple case of one-dimensional periodic
lattice in Section~\ref{sec1d}. In Section~\ref{sec2d} we calculated the
mean value and the variance of the number of bonds for the square lattice
using the results of Section~\ref{sec1d}. In the following
Section~\ref{secany} we genelarize our results for the $d$-dimensional
hypercubic lattice. We conclude our paper with the discussion of results
in Section~\ref{analyze}.

\section{One-dimensional lattice}
\label{sec1d}

\subsection{s-sampling}

Let us consider an one-dimensional lattice with $L$ sites and
periodic boundary (i.e. ring). 

We are interested in the probability $P(b;s,L)$ to have $b$ bonds between
$s$ occupied sites. The number of occupied sites $s$, the number of bonds
between occupied sites $b$ and the number of clusters of occupied sites
$k$ are connected by relation
\begin{equation}
k=s-b.
\label{ksb}
\end{equation}
Then, the probability could be easily written as
\begin{equation}
P(b;s,L)=\frac{LC_{s-1}^{k-1}C_{L-s-1}^{k-1}}{kC_L^s},
\label{PbsL}
\end{equation}
where combinatorial coefficients $C_{s-1}^{k-1}$ is the number of
dispositions of $s$ sites among
the $k$ clusters, $C_{L-s-1}^{k-1}$ - the same for $L-s$ empty sites,
$C_L^s$ - number of configurations with the $s$ occupied sites from $L$
possibilities, and, finally, $L/k$ - the number of positions of $k$
clusters on $L$ sites. The probability are normalized
\begin{equation}
\sum_{b=0}^{s-1} P(b;s,L)=
\sum_{b=0}^{s-1}\frac{LC_{s-1}^{s-b-1}C_{L-s-1}^{s-b-1}}{(s-b)C_L^s}=1.
\label{normP}
\end{equation}
Now, the mean value of bonds $\left<b\right>$ are
\begin{equation}
\left<b\right>=\sum_{b=0}^{s-1}bP(b;s,L)=\frac{s(s-1)}{L-1}
\label{bavs}
\end{equation}
and the mean density of bonds are given by $L>>1$
\begin{equation}
n_b=\left<b\right>/L\approx
n_s^2-n_s(1-n_s)/L,
\label{nbs}
\end{equation}
where $n_s=s/L$ is the density of occupied sites.

The variation of the number of bonds $\Delta
b=\left<b^2\right>-\left<b\right>^2$ could be calculated
in the same manner
\begin{equation}
\Delta b=\sum_{b=0}^{s-1}b^2 P(b;s,L)-\left(\sum_{b=0}^{s-1}bP(b;s,L)\right)^2=
\frac{s(s-1)(L-s)(L-s-1)}{(L-1)^2(L-2)}\approx \frac{s^2(L-s)^2}{L^3}
\label{Dbs}
\end{equation} 
and the variance of the density of bonds $\Delta n_b$ are equal to
\begin{equation}
\Delta n_b\approx n_s^2(1-n_s)^2-n_s(1-n_s)(4n_s^2-4n_s+1)/L
\label{Dnbs}
\end{equation}
for $L>>1$.

\subsection{p-sampling}

The probability distributions of the some quantity $A$ in the case of
s-sampling $P(A;s)$ and those of the case of p-sampling $P(A;p)$ are
connected by

\begin{equation}
P(A;p)=\sum_{s=0}^LP(A;s)C_L^sp^s(1-p)^{L-s},
\label{p-s-relation}
\end{equation}
therefore the probability to obtain exactly $b$ bonds between two sites
occupied with the (independent) probability $p$ on the lattice of
size $L$ are
\begin{equation}
P(b;p,L)=\sum_{s=0}^L \left( C_L^sp^s(1-p)^{L-s} 
\frac{LC_{s-1}^{k-1}C_{L-s-1}^{k-1}}{kC_L^s}\right)
\label{PbpL}
\end{equation}
with $k$ given by eq.~(\ref{ksb}) and the mean value of the bonds $\left<b\right>'$
in the case of p-sampling will be
given by
\begin{equation}
\left<b\right>'=\sum_{b=0}^{s-1} bP(b;p,L)\equiv
\sum_{s=0}^L\left( C_L^sp^s(1-p)^{L-s}\left<b\right> \right)=Lp^2,
\label{bavp}        
\end{equation}
(where prime sign $'$ denotes the p-sampling averages) and the
corresponding variation $\Delta b'$ are given by
\begin{equation}
\Delta b'=\sum_{s=0}^L\left( C_L^sp^s(1-p)^{L-s}\; 
\left(\Delta b\+\left<b\right>^2\right) \right) 
- \left<b\right>'^2=Lp^2(1-p)(1+3p)
\label{Dbp}
\end{equation}
for $L \ge 4$. 

We have to stress here, that the mean density of the number of bonds
$n_p=p^2$ and their variance $\Delta n_p=p^2(1-p)(1+3p)$ do not contain
any finite size dependences in contrast with the case of s-sampling
behaviour in (\ref{nbs}) and {\ref{Dnbs}).

\section{Square lattice}
\label{sec2d}

Let us calculate the mean value of the number of bonds and it's variation
for the square lattice with periodic boundary conditions.   

\subsection{p-sampling}
\label{g-2d}

On the square lattice with periodic boundaries, there are $L^2$ vertical
and $L^2$ horizontal bonds. The number of bonds in each of $L$ columns and
each of $L$ rows are the random variables. Their mean values
$\left<b\right>'_i$ are given by (\ref{bavp}) and their variances by
(\ref{Dbp}). Thus, the mean value of the number of bonds on 2d lattice are

\begin{equation}
\left<b\right>'_{d=2}=\sum_{i=1}^{2L} \left<b\right>_i'=2Lp^2.
\label{bavp2cr}
\end{equation}
and their variance are given by~\cite{Feller}
\begin{equation}
\Delta b'_{d=2} =\sum_{i=1}^{2L}  \Delta b' 
+ 2 \sum_{i,j}^{2L} {\rm Cov} \left(\left<b\right>_i',\left<b\right>_j'\right)
\label{bdavp2cr}
\end{equation}
where $i<j$. 

It is clear that the column-column (row-row) covariances are equal to
zero, and the column-row covariances are the same for all columns and
rows, and 
\begin{equation} 
\Delta b'_{d=2} = 2L^2p^2(1-p)(1+3p)+2L^2 {\rm
Cov} \left(\left<b\right>_{column}',\left<b\right>_{row}'\right).
\label{bavp2cr1} 
\end{equation}
 
We have to compute now the column-row covariance ${\rm Cov}
\left(\left<b\right>_{column}',\left<b\right>_{row}'\right)$

\begin{equation} 
{\rm Cov} \left(\left<b\right>_{column}',\left<b\right>_{row}'\right)=
\left<\left<b\right>'_{column}\left<b\right>'_{row}\right>
-\left<b\right>'^2.  
\end{equation} 

The first average are splitted into two parts. The first one, corresponds
to all cases of distribution of occupied sites on columns and rows with
the constraint that the intersection of row and column $x^*$are occupied.
The second one, corresponds to the states with the unoccupied intersection
site.

\begin{equation}
\left<\left<b\right>'_{column}\left<b\right>'_{row}\right>=
p\left<b|x^*=occupied\right>^2+(1-p)\left<b|x^*=unoccupied\right>^2
\label{split}
\end{equation}

We omit here the algebra leading to the final expression for the variation
of the number of sites on the two-dimensional lattice with the given
probability $p$ for the site to be occupied

 \begin{equation}
\Delta b'_{d=2}=2L^2p^2(1-p)(1+7p).
\end{equation}
Details will be published elsewhere \cite{Vas1}.

\subsection{s-sampling}

We calculate also the mean value of the number of bonds on the square
lattice with the linear size $L$ and with the fixed number $s$ of occupied
sites

\begin{equation}
\left<b\right>_{d=2}=\frac{2s(s-1)}{L^2-1}
\label{2db}
\end{equation}
and the variance are
\begin{equation}
\Delta b_{d=2}=
\frac{s(s-1)(L^2-s)(L^2-s-1)(2L^2-10)}{(L^2-1)^2(L^2-2)(L^2-3)}.
\label{2dvarb}
\end{equation}

\section{Hypercubic lattices}
\label{secany}

The mean value of the number of bonds on the $d$-dimensional hypercubic
lattice with the probability $p$ for the site to be occupied are
\begin{equation}
\left<b\right>'_d=dL^dp^2.
\label{anybp}
\end{equation}

The variance could be calculated in the same manner as in the two
dimensional $d=2$ case described above in the Section~\ref{g-2d}

\begin{equation}
\Delta b'_d=dL^dp^2(1-p)(1+\left(4d-1)p\right).
\label{anydbp}
\end{equation}

The explicit mean value and the variance of the number of bonds for the
case of the fixed number $s$ of occupied sites could be calculated using
(\ref{anybp}) and (\ref{anydbp}) with the proposition that the mean value
are quadratic in $s$. The result are

\begin{equation} 
\left<b\right>_{d}=\frac{ds(s-1)}{L^d-1} 
\label{anybs}
\end{equation} 

and 

\begin{equation}
\Delta b_{d}=
\frac{ds(s-1)(L^d-s)(L^d-s-1)(L^d-2d-1)}{(L^d-1)^2(L^d-2)(L^d-3)}
\label{anydbs} 
\end{equation}
for $L>2$.

The expressions (\ref{anybp}-ref{anydbs}) we checked by direct enumeration
in $d=2$ for $L=4,5$ and in $d=3$ for $L=3$~\cite{Vas1}.

\section{Discussion} 
\label{analyze}

We do not find any evidence for the absense of self-averaging
\cite{DerHil,BH-book} in the number of bonds. In the case of
p-sampling the normalized variance 

\begin{equation} 
R_{b_d'}=\frac{\Delta b'_d}{\left<b\right>'^2_d}=
\frac{(1-p)(1+(4d-1)p)}{p^2L^d}\propto L^{-d} 
\end{equation} 

inverse proportional to the lattice volume. Therefore the number of
bonds are the strong averaging quantity~\cite{WisDom,AHW}.

The normalized variance in the case of s-sampling

\begin{equation}
R_{b_d}=\frac{\Delta b_d}{\left<b\right>_d^2}=
\propto \frac{(1-n_s)^2}{dL^{-d}n_s(n_s-1/L^d)}
\end{equation}

are nonmonotonic in $L$ and the strong self-averaging take place for
$L>\left(1/n_s\right)^{1/d}$.

The relative variance for two cases

\begin{equation} 
\frac{\Delta b'_d}{\Delta b_d}=
\frac{p^2(1+(4d-1)p)(1-p)}{\left( n_s(1-n_s) \right)^2}
\label{ratio} 
\end{equation} 
shows that in the limit of $p\rightarrow1$ and $n_s\rightarrow1$ the
ratio diverges. 

It is possible to give the physical interpretation to the last expression.
Let us place Ising spin at each occupied site.  Then, the energy $E_0$ at
zero temperature will be equal to the number of bonds betwen the sites
occupied by spins. The magnetization $M_0$ are equal to the number of
occupied sites. Then, the ratio (\ref{ratio}) are equal to the ratio
$N_p/N_s$ of the number of samples ($N_p$ for the p-sampling and $N_s$ for
the s-smapling) over which we have to average energy $E_0$ in order to get
the same dispersion $\sigma E_0=\sqrt{\Delta b/(N_x-1)}$ of the energy
$E_0$, where $N_x$ are either $N_p$ or $N_s$.

Finally, we could propose the third way to generate the samples, {\it
sb-sampling}: get random samples with the fixed number of sites $s$ and
choose among them only those which contain the "right" number of bonds
$b=ds^2$. The results of Monte Carlo simulations for site diluted Ising
model on square lattice had show \cite{Vas1} that the dispersions of
termodynamic quantaties like energy, specific heat and magnetic
susceptibility are quite smaller for the proposed sb-sampling in
comparison with both, s-sampling and p-sampling at all temperatures.

\section{Acknowledgments}

Authors are thankfull to K.~Binder for the kind discussion and suggestion
to write this manuscript. The work is partially supported by grants from
RFBR (99-02-18412), NWO and INTAS. O.A.V. is grateful to Landau stipendium
committee (Forshungzentrum/KFA, J\"ulich) for support.


\begin{references} 

\bibitem{RD-12} L.N. Shchur, {\it Incipient Spanning Clusters in Square
and Cubic Percolation}, in Springer Proceedings in Physics, "Computer
Simulation Studies in Condensed Matter Physics XII", Eds. D.P. Landau,
S.P. Lewis, and H.B. Sch\'h"uttler, (Springer-Verlag, Berlin, 2000)

\bibitem{GrifLeb} R.B. Griffiths and J.L. Lebowitz, J. Math. Phys. {\bf 9}
(1968) 1284

\bibitem{Feller} W. Feller, {\it An introduction to probability theory and
its applications} (John Wiley \& Sons, New York, 1970)

\bibitem{Vas1} O.A. Vasilyev, to be published.

\bibitem{DerHil} B. Derrida and H. Hilhorst, J. Phys. C: Solid State
Phys. {\bf 14} (1981) L539

\bibitem{BH-book} K. Binder, D.W. Heermann, {\it Monte Carlo Simulations
in Statistical Physics} (Springer-Verlag, Berlin, 1997)


\bibitem{WisDom} S. Wiseman and E. Domany, Phys. Rev. E {\bf 52}
(1995) 3469;  Phys. Rev. Lett. {\bf 81} (1998) 22; Phys. Rev. E {\bf 58}
(1998) 2938

\bibitem{AHW}
A. Aharony and A. B. Harris, Phys. Rev. Lett. {\bf 77} (1996) 3700;
A. Aharony, A. B. Harris and S. Wiseman, Phys. Rev. Lett. {\bf 81}
(1998) 252

\end{references}
\end{document}